\documentclass[prb,twocolumn,amssymb,amsfonts,superscriptaddress]{revtex4}
\usepackage{epsfig}
\usepackage{amsmath}

\begin{document}

\title{Electrostatic traps for dipolar excitons}

\author{Ronen Rapaport, Gang Chen, Steve Simon, Oleg Mitrofanov, Loren Pfeiffer, and P. M. Platzman}
 \affiliation{Bell Laboratories, Lucent Technologies,
600 Mountain Avenue, Murray Hill, New Jersey 07974}

\begin{abstract}

We consider the design of two-dimensional electrostatic traps for
dipolar indirect excitons. We show that the excitons dipole-dipole
interaction, combined with the in-plane electric fields that arise
due to the trap geometry, constrain the maximal density and
lifetime of trapped excitons. We derive an analytic estimate of
these values and determine their dependence on the trap geometry,
thus suggesting the optimal design for high density trapping as a
route for observing excitonic Bose-Einstein condensation.
\end{abstract}

\maketitle

For many years, excitons in semiconductors had been predicted to
undergo a phase transition at high enough densities and low enough
temperatures to form a Bose-Einstein condensate
(BEC)\cite{ButovJPCM2004,SnokeScience2002,Snokebook}, similar to
BEC of atomic gases, already observed a decade
ago\cite{WienmannKetterleRMP2002}. This is expected to happen due
to the predicted bosonic nature of excitons at densities that
still disguise the fermionic nature of their constituents, i.e.,
the electron and the hole. A few major obstacles have, however,
prevented a clear observation of an exciton BEC phase until this
day, even though the typical predicted transition temperature is
of the order of a kelvin, much hotter than its atomic counterpart,
and is available in many labs. Maybe the most crucial obstacle to
excitonic BEC is the short exciton intrinsic radiative lifetime
(of the order of hundred picosecond) due to electron-hole
recombination, which limits the time available for exciton
thermalization.  Since the initial state of the exciton after
optical excitation is out of equilibrium, and full thermalization
with the lattice becomes more difficult at low lattice
temperatures, the thermalization time turns out to be longer than
the intrinsic exciton lifetime. Thus, the temperature of the
exciton gas may not reach the required transition temperature to
the condensed state, although the lattice temperature may in fact
be well below that temperature. In recent years, a promising way
to overcome the lifetime issue has emerged. The exciton lifetime
can be considerably increased by spatially separating the electron
and the hole. This is usually achieved by utilizing a double
quantum well (DQW) system \cite{AlexandrouPRB1990,NegoitaPRB1999}.
The resulting excitons are constructed from electrons in one layer
and holes in the other and are known as ``spatially indirect
excitons". This trick can increase the exciton lifetime by many
orders of magnitude (from less than a nanosecond to tens of
microseconds) \cite{SnokeCondmat2005} while only slightly reducing
the exciton binding energy (due to the 3D nature of the coulomb
interaction).

It seems that by utilizing these indirect excitons, the major
obstacle to BEC has been removed. However, a new problem arises.
The indirect excitons are dipolar in nature since they all carry a
permanent dipole moment due to the charge separation of the
electron and hole in the growth direction, perpendicular to the QW
planes. All the dipoles are aligned in the same direction. As a
result, there is a strong repulsive dipole-dipole interaction
between all excitons. On one hand, this repulsive interaction has
an additional advantage, since it prevents further binding of
excitons into larger complexes, such as biexcitons or
electron-hole droplets. On the other hand, the strong repulsive
interaction will also tend to spread the free exciton gas and to
quickly reduce its density. Thus it is difficult to maintain a
very dense gas of free dipolar excitons over long times due to
this driven expansion. Recent experiments probing the dynamics of
a dense excitonic gas indeed show a very fast expansion of the
dense cloud over a short period of time followed by a much slower
expansion of the dilute cloud \cite{SnokeCondmat2005}. In a
different work we show how this behavior can be quantitatively
explained by an initial driven fast expansion that transforms into
diffusive expansion when the density drops
\cite{RapaportUnpublished}. One can show that at densities and
temperatures required for observing excitonic BEC, the dipolar
exciton gas expansion will always be initially driven outwards by
the strong repulsive dipole-dipole interactions, quickly reducing
its density. One possible solution for this problem is to create a
homogeneous distribution of such excitons over the whole sample.
This will indeed eliminate the fast expansion by eliminating the
density gradients. Such a solution will, however, require a lot of
excitation power and will end up heating the sample. One can think
of a more elegant solution in which the dipolar excitons are
trapped in an external potential, preventing them from expanding.
Such methods have been extremely successful in trapping and
cooling atoms, leading to their condensation
\cite{WienmannKetterleRMP2002}. Since the excitons are already
confined in one dimension (by quantum confinement of the quantum
well), one needs to take care of only the in-plane confinement.
One possibility of confining the excitons in the plane is by the
use of applied localized stress to change the local band energies
\cite{SnokeCondmat2004}. In this paper we discuss an alternate
scheme of circular, two-dimensional electrostatic traps which trap
the dipolar excitons in a well-defined space. This trapping occurs
via the interaction of the exciton's permanent dipole with a
non-uniform electric field. It was shown before that spatially
indirect excitons can be transported \cite{HagnAPL1995} and
trapped in a one-dimensional periodic way \cite{ZimmermannAPL1998}
using spatially varying electric fields. The electrostatic
trapping method allows trapping of dipolar excitons in a wide
range of trap sizes, and can also enable a fast, dynamic control
of the trapped excitons by electrical modulation of the trap depth
and shape, thus allowing, for example, evaporative cooling of the
exciton gas.

Here we consider the limitations on the excitonic density and
effective lifetime of such a trapping method. We show that there
is a minimum required vertical applied electric field to get the
trapping energy larger than the dipole-dipole repulsion energy, in
order to prevent the trapped excitons from escaping. However, in
general, applying a vertical field will also result in a radial
electric field depending on the geometry of the trap. This radial
electric field will in turn cause exciton ionization at the trap
boundaries and reduce the effective trapping lifetime. We then
derive an analytic estimation of the maximal density and lifetime
of trapped excitons. This analysis gives guidelines as well as
constraints for optimal design of such dipolar traps.

\vspace*{0cm}
\begin{figure}[htb]
\begin{center}
\includegraphics[scale=0.4]{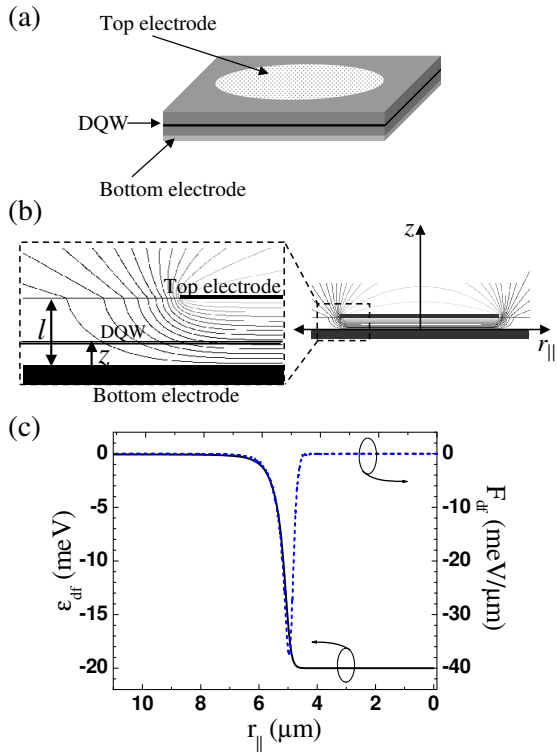}
\caption{(a)An illustration of a suggested dipolar trap design.
(b) Equipotential lines of a circular trap with $R/l=10$. (c)
Calculated confining energy for dipolar excitons
($\varepsilon_{df}$, solid line) and force ($F_{df}$, dashed line)
for a circular trap with $R=5 \mu m$, $l=0.5 \mu m$, a potential
difference $\triangle \phi_0$ of $1V$ and a dipole length
$z_0=100$\AA.} \label{figure1}
\end{center}
\end{figure}

The physical idea of an electrostatic trap is straightforward.
Consider a geometry where a small circular, optically
semi-transparent metallic gate with a radius $R$ is placed on top
of the DQW sample. The sample substrate is made conductive to
serve as the ground electrode, as illustrated in
Fig.~\ref{figure1}a. The sample thickness, from the bottom to the
top electrode, is denoted by $l$. The DQW plane is perpendicular
to the growth direction, $(\hat{z})$ and its vertical position in
the sample is given by $z$, measured from the bottom electrode.
The indirect excitons will tend to stay in the region under the
gate contact when a sufficient voltage is applied between it and
the substrate. This trapping effect is due to the fact that the
indirect excitons are dipolar and thus they are high field
seekers, gaining an additional negative energy term coming from
the dipole-field interaction:
\begin{equation}
\varepsilon_{df} = \vec{d}_{X} \cdot
\vec{E}(r_{\|},z)=d_XE_z(r_{\|},z), \label{dfenergy}
\end{equation}
where $\vec{d}_{X}=-ez_0\hat{z}$ is the exciton dipole moment,
$\vec{E}=-\nabla\phi$ is the applied electric field, and $z_0$ is
the effective separation between the electron and the hole and is
equal to the separation between the centers of the two quantum
wells to a good approximation. Fig.~\ref{figure1}b shows a
cross-section of a circular trap with equipotential lines
illustrated, where $R/l = 10$. Fig.~\ref{figure1}c depicts the
confining energy, $\varepsilon_{df}$, as a function of the radial
position of the excitons in the trap, $r_{\|}$, for $R=5 \mu m$,
$l=0.5 \mu m$, $z_0=100$\AA and a potential difference
$\triangle\phi_0$ of $1V$. The trapping energy at the center of
the trap (where $E_z^{center} \equiv \triangle\phi_0/l$) is given
by:
\begin{equation}
\varepsilon_{df}(r_{\|}=0,z)=-ez_0\triangle\phi_0 /l.\label{trapenergy}
\end{equation}
Since $E_z$ varies significantly only over a range $\Delta r
\approx l$ near $r_{\parallel}=R$, then for $R\gg l$, the excitons
experience the trapping dipole force just near the sharp
boundaries, given by:
\begin{equation}
F_{df}(r_{\|},z) =
d_X\frac{\partial{E_z}}{\partial{r_{\|}}}\hat{r_{\|}}
\label{dfforce}
\end{equation}
as is seen in Fig.~\ref{figure1}c.

Ideally, for $R\gg l$, such a trap behaves like a "pool" of free
moving excitons, subject to perfectly reflecting boundary
conditions at the edges (one can get a non-flat potential well for
$R\sim l)$. One has to compare the trapping energy
$\varepsilon_{df}$, to the dipole-dipole repulsion energy,
$\varepsilon_{dd}$ and to $kT$. As long as
$\varepsilon_{df}>\varepsilon_{dd},kT$, the excitons will be
confined within the boundaries of the trap, being reflected from
the walls by the trap's dipole force. If, however, the opposite
condition arises, the excitons will "spill over" the trap due to
the internal repulsive force. The dipole-dipole repulsion is given
by:
\begin{equation}
\varepsilon_{dd}=\frac{4\pi e^2z_0}{\epsilon}n_X, \label{ddenergy}
\end{equation}
where $\epsilon$ is the background dielectric constant. The
condition for trapping, \begin{equation}
\mid\varepsilon_{df}/\varepsilon_{dd}|
> \alpha
\end{equation}
where $\alpha$ where $\alpha$ is a parameter of order 1,
determined by how much ``residual" trapping energy is required in
the experiment.  This condition then yields the maximal trapped
exciton density:
\begin{equation}
n_X^{max}= \frac{\epsilon E_z^{center}}{4\pi e\alpha},
\label{trapcond1}
\end{equation}
Here, we have neglected the thermal energy $kT$ since it is always
much smaller than $\varepsilon_{dd}$ at temperatures and densities
relevant for excitonic BEC. This inequality actually leads to some
interesting consequences and is discussed elsewhere
\cite{RapaportUnpublished}.

\begin{figure}[htb]
\begin{center}
\includegraphics[scale=0.4]{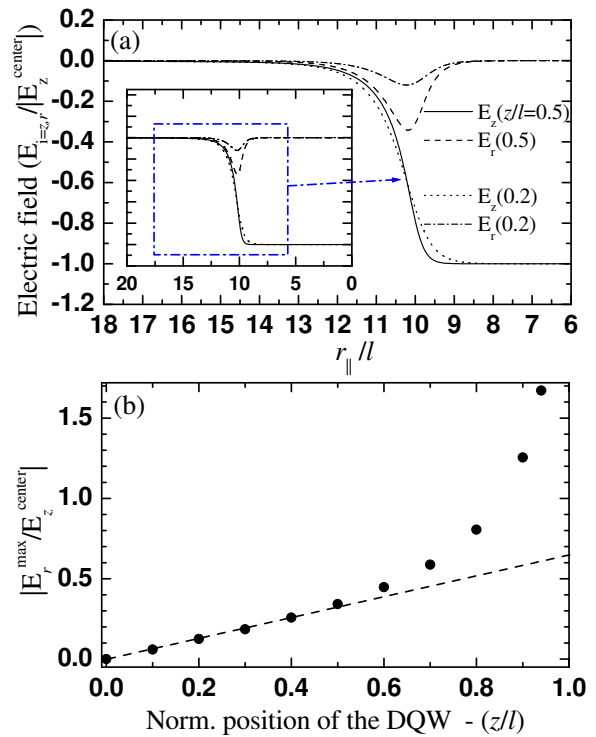}
\caption{(a) Vertical ($E_z$) and radial ($E_r$) components of the
electric field in a circular trap as a function of $r_{||}$ for
two different vertical positions ($z/l=0.2,0.5$) of the DQW
structure. The inset shows the whole range of radial positions
from the center of the trap. (b) The ratio of the maximal value of
the radial component of the electric field, $E_r^{max}$, to the
vertical component of the field at the trap center,
$E_z^{center}$, for various vertical position $z/l$ values
(circles). The dashed line is a linear fit to the range $0\leq z/l
< 5$}\label{figure2}
\end{center}
\end{figure}

Eq.~\ref{trapcond1} seems to suggest that an arbitrary density of
excitons can be trapped, depending only on $E_z^{center}$ and thus
on the external potential difference. Unfortunately, the picture
given above is too simplified. While the electric field inside the
trap and away from the edge is always aligned with the dipoles
(i.e. $\vec{E}\cong E_z \hat{z}$), at the boundary of the trap the
radial component of the electric field, $E_r$, can be appreciable,
depending on $z$. As $z$ increase (moving the quantum well away
from the bottom electrode), the magnitude of $E_r$ increases
compared to $E_z$, as is shown in Fig.~\ref{figure2}a for two
different exemplary $z/l$ positions. One can extract the ratio of
the maximal radial component of the electric field, $E_r^{max}$ to
the vertical component at the center of the trap, $E_z^{center}$.
This is plotted in Fig.~\ref{figure2}b. For $z/l<1/2$, this ratio
can be well approximated with a linear dependence,
\begin{equation}
\left|\frac{E_r^{max}}{E_z^{center}}\right|\cong\beta z/l,
\label{Eratio}
\end{equation}
with $\beta=0.625$.

\vspace*{0cm}

\vspace*{0cm}
\begin{figure}[htb]
\begin{center}
\includegraphics[scale=0.6]{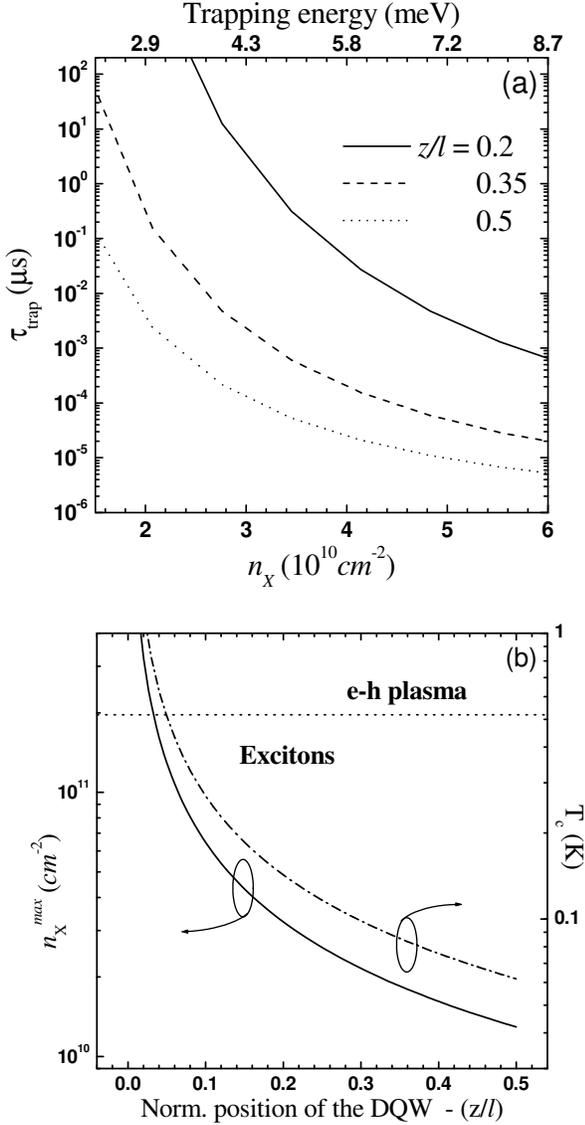}
\caption{(a) Effective trap lifetime as a function of the trapped
density $n_X$ and the trap energy $\varepsilon_{df}$ for various
vertical positions of the DQW structure in the trap, $z/l$. Here
 $R/l=50$, $\varepsilon_X=5meV$, and $\alpha=1.2$.
(b) The calculated maximal density (solid line) of trapped dipolar
excitons (with same parameters as in (a)), and the expected BEC
transition temperature (dashed-dot line), as a function of $z/l$.
The horizontal line marks the approximated Mott
density.}\label{figure3}
\end{center}
\end{figure}

The radial field will tend to pull the electron and the hole into
opposite directions. Classically, if the electrostatic energy due
to the in-plane external field is larger than the exciton binding
energy, the exciton will be ionized. This happens when
$e|E_r|a_X\approx\varepsilon_X$, where $\varepsilon_X$ is the
dipolar exciton binding energy and $a_X$ is its in-plane radius.
Quantum mechanically however, one expects that ionization of
excitons becomes significant at much smaller values of in-plane
field, due to tunnelling of the electron and hole through the
``hill" in the coulomb potential that binds the exciton
\cite{ZimmermannPRB1997}. Such tunnelling will break some of the
excitons on the edge of the trap, thus decreasing the total
density of excitons with time and giving rise to an effective trap
lifetime. In other words, the quantum ionization process
effectively makes the perfectly reflecting boundaries of the trap
to become partly absorbing.

The ionization rate of a 2D exciton in its ground state subject to
a one-dimensional electric field has been calculated in
Ref.~\onlinecite{MillerPRB1985}. We define the ratio between the
exciton binding energy and the typical electrostatic energy the
exciton experiences due to the in-plane field as:
\begin{equation}
\gamma=\varepsilon_X/(e|E_r|a_X), \label{gammadef}
\end{equation}
Where the local field correction due to the induced in-plane
polarization of the excitons can be neglected \cite{localfield}.
Since $E_r$ changes over a length scale $\sim l\gg a_X$, we follow
Ref.~\onlinecite{MillerPRB1985} to get the ionization rate of the
exciton at the trap boundary:
\begin{equation}
\Gamma_{ion}=\frac{32\varepsilon_X}{\sqrt{\pi}\hbar}\sqrt{\gamma}
e^{-(8\gamma/3)}. \label{ionization}
\end{equation}
Assuming a homogenous distribution of excitons in the trap
($N_X=n_X\pi R^2$), the number of excitons within a distance $l$
from the edge is given by $n_X2\pi R\triangle R\approx n_X2\pi
Rl$. The depletion of the exciton density in the trap can be
described by a simple rate equation:
\begin{equation}
\frac{dn_X}{dt}=-\frac{2l}{R}\Gamma_{ion}n_X=-\frac{n_X}{\tau_{trap}}
\label{deprate}
\end{equation}
where we have defined an effective trap lifetime,
$1/\tau_{trap}=(2l/R)\Gamma_{ion}$ and assumed
$\tau_{trap}<<\tau_X$ ($\tau_X$ being the exciton intrinsic
lifetime).

The constraints on the trap performance can already be seen: due
to the dipole-dipole interaction $\varepsilon_{dd}$, there is a
minimum required vertical field in order to get the trapping
energy $\varepsilon_{df}$ larger than $\varepsilon_{dd}$ for a
given required trap density (Eq.~\ref{trapcond1}). This
requirement will introduce a radial field due to the geometrical
relationship between $E_z$ and $E_r$ given by Eq.~\ref{Eratio}.
The radial field will in turn increase the ionization rate at the
trap boundaries and will reduce the effective trap lifetime
(Eqs.~\ref{deprate} and ~\ref{ionization}).

By combining Eqs.~(\ref{trapcond1})-~(\ref{deprate}) one can get
the effective trap lifetime as a function of the trap energy (and
hence the maximal trapped exciton density):

\begin{equation}
\frac{1}{\tau_{trap}}=\frac{32}{\pi}\sqrt{\frac{\epsilon
l^2(l/z)\varepsilon_X^3}{\hbar^2R^2\alpha\beta
e^2a_Xn_X}}\exp{\left(\frac{-2(l/z)\epsilon\varepsilon_X}{3\pi\alpha\beta
e^2a_Xn_X}\right)}. \label{tautrap}
\end{equation}

This is plotted in Fig.~\ref{figure3}a for various $z/l$ values,
for a trap with $R/l=50$, $\varepsilon_X=5meV$, and setting
$\alpha$ to be 1.2. As the exciton density increases, there is a
strong reduction of $\tau_{trap}$. This will dramatically reduce
the time available for exciton thermalization even if their
intrinsic lifetime is very long.

Inverting the previous argument yields a bound on the maximal
$E_r^{max}$ (through the unitless parameter $\gamma$), for a
minimum desired effective trap lifetime, $\tau_{trap}$ which we
should choose to be long enough for efficient exciton
thermalization for BEC. With such a requirement,
Eq.~(\ref{ionization}) and ~(\ref{deprate}) give:
\begin{equation}
\gamma - \frac{3}{16}\log_e\gamma =
\frac{3}{8}\log_e\left(\frac{64l}{\sqrt{\pi}R}\frac{\tau_{trap}}{\tau_{gs}}\right),
\label{trapcond2}
\end{equation}
where we have defined $\tau_{gs}=\hbar /\varepsilon_X$. This bound
\cite{3Dlimit} yields an upper limit on $E_z^{center}$ through the
geometrical relation of Eq.~(\ref{Eratio}). The upper limit on
$E_z^{center}$ then sets a limit on $n_X^{max}$, due to the
competition of $\varepsilon_{df}$ and $\varepsilon_{dd}$, as
reflected in Eq.~(\ref{trapcond1}).

Combining Eqs.~(\ref{Eratio}) and (\ref{trapcond1}) results in a
compact expression for estimating the maximal density of excitons
that can be trapped, depending on the various requirements and
trap design:
\begin{equation}
n_X^{max}=\frac{(\epsilon\varepsilon_X)^2}{4\pi
e^4}\frac{(l/z)}{\alpha\beta\gamma}. \label{trapcond3}
\end{equation}
Fig.~\ref{figure3}b shows the maximal trapped density as a
function of $(z/l)$, for $\tau_{trap}=1\mu s$ and with the same
trap parameters as before. Note that $\gamma$ depends
logarithmically on $\tau_{trap}$, hence relaxing the trapping time
constraint by orders of magnitude will result only in a small
increase of $n_X^{max}$. The predicted BEC transition temperature,
$T_c$ for the exciton gas in a circular trap with $R=25\mu m$,
given by \cite{IvanovEurophys2002} $T_c=\pi\hbar^2
n_X^{max}/(2k_BM_X\log_e(n_X^{max}\pi R^2/4))$, is also plotted.
As the DQW gets further away from the bottom electrode, there is a
strong reduction of $n_X^{max}$ due to the increased in-plane
electric field. Thus, this is an important issue in a design of a
trap. The guideline for designing a trap for high density dipolar
exciton gas is then to minimize $z/l$ as much as possible ($z/l$
should be smaller than $0.25$ for an exciton density larger than
$10^{10} cm^{-2}$).

There are few other possible ways of getting around this
complication: (a) one can design a trap with symmetric top and
bottom electrodes. This will eliminate in-plane fields exactly at
$z/l=1/2$. However, constructing such a trap is much more
difficult from a fabrication point of view. (b) it is possible to
use a deep trap and continuously pump the trap to achieve a higher
steady-state density to compensate for the fast tunnelling time.
This will, unfortunately, tend to heat up the trap while the time
available for excitons to thermalize will be significantly shorter
due to the fast ionization, even for intrinsically long lifetime
excitons, as can be seen from Fig.~\ref{figure3}a. (c) finally, in
order to keep the DQW structure close to the center ($z/l\simeq
1/2$), but minimize the in-plane field ionization problem, a trap
can be designed where a doped QW layer is inserted just below the
DQW structure. This will greatly reduce the $E_r$ component at the
DQW due to its vicinity to a charged, metallic-like layer.

In summary, we have analyzed the constraints on the design of
electrostatic traps for dipolar excitons and derived expressions
relating the trapping lifetime and the trap design parameters to
the maximal density of excitons that can be trapped and to their
corresponding expected BEC transition temperature. We show that it
is feasible to construct dipolar traps that will trap excitons
with high enough densities and for long times for a possible
observation of excitonic BEC. Strong experimental evidence for
high density exciton trapping has already been observed in our
lab, utilizing similar trap designs as discussed above, and will
be presented in a separate paper.

After the completion of this manuscript, we were informed of a
related ongoing work by L. V. Butov and co-workers.


\begin{thebibliography}{14}
\bibitem{localfield} The local field correction due to the induced in-plane dipole of the excitons can be approximated by
$E_r^{ind}\approx (4n_Xz_0/l)E_r\ll E_r$ and thus can be
neglected.
\bibitem{3Dlimit} Using a three-dimensional excitonic model leads to
$\gamma$ values larger by $\sim 2$, thus reducing the maximal
estimated density by the same factor.
\end{thebibliography}
\end{document}